\documentclass[11pt]{article}

\usepackage[T1]{fontenc}
\usepackage[left=2.5cm, right=2.5cm, top = 2.5cm, bottom = 4cm]{geometry}
\usepackage{xcolor}
\usepackage{amsmath}
\usepackage{amssymb}
\usepackage{braket}
\usepackage{graphicx}
\usepackage[sorting=none]{biblatex}
\usepackage{hyperref}
\hypersetup{
    colorlinks = true,
    linkcolor = blue,
    anchorcolor = blue,
    citecolor = blue,
    filecolor = blue,
    urlcolor = blue
}

\newcommand{\op}[1]{\hat{#1}}
\newcommand{\mi}{\mathrm{i}}
\newcommand{\lpar}[1]{\left(#1\right)}
\newcommand{\lspar}[1]{\left[#1\right]}
\newcommand{\lbr}[1]{\left\{#1\right\}}

\title{Intrinsic decoherence for the displaced harmonic oscillator}
\author{Alejandro R. Urz\'ua\footnote{email: arurz@inaoep.mx} and H\'ector M. Moya-Cessa\footnote{email:hmmc@inaoep.mx}\\
    Coordinaci\'on de \'Optica\\ Instituto Nacional de Astrof\'isica, \'Optica y Electr\'onica\\
    Luis Enrique Erro \#1, 72840, Tonantzintla, Puebla}
\date{\today}

\begin{document}

\maketitle

\begin{abstract}
By using the complete solution of the Milburn equation (beyond the Lindblad form that it is generally used) that describes intrinsic decoherence, we study the decaying dynamics of a displaced harmonic oscillator. We calculate the expectation values of position quadrature, and the number operator in  initial coherent  and squeezed states.
\end{abstract}

\section{Introduction}
Being decoherence the main enemy of the nonclassical properties that may be generated in quantum mechanical systems, it is of great interest its study to learn how the interesting properties that  arise in such systems may be maintained. Some years ago Milburn \cite{Milburn1991} proposed a modification of the Schr\"odinger equation that accounts for (intrinsic) decoherence. By assuming that the system evolves by a random sequence of unitary phase changes on sufficiently small time scales Milburn produced a Lindblad equation with the Hamiltonian as the relevant operator. Since then, several authors have studied the decay of coherences for different systems, for instance it has been shown that in the atom-field interaction such loss of coherences prevents the revivals to occur for the atomic population inversion \cite{MoyaCessa1993}.  Yang {\it et al.} \cite{Yang2017} determined the performance of quantum Fisher information of the two-qutrit isotropic Heisenberg $XY$ chain subject to decoherence; Mohamed {\it et al.} \cite{Mohamed2021} analyzed the robustness of quantum correlations of the nearest neighbour and the next-to-neighbour qubits in an intrinsic noise model describing the dynamics of the decoherence for a system formed by three-qubit Heisenberg $XY$ chain.  Zheng and Zhang \cite{Zheng2017} applied Milburn's scheme to study the entanglement in the Jaynes Cummings model, where a pair of atoms undergo  Heisenberg type interactions; He and Chao \cite{He2021} studied the coherence dynamics of two atoms in a Kerr-like medium; Muthuganesan and Chandrasekar \cite{Muthuganesan2021} applied intrinsic decoherence when studying an exactly solvable model of two interacting spin-$\frac{1}{2}$ qubits described by the Heisenberg anisotropic interaction. Chlih \textit{et al.} \cite{Chlih2021} used the intrinsic decoherence to study a variety of initial states, where they obtain the temporal evolution of quantum correlations in a two-qubit XXZ Heisenberg spin chain model subject to a Dzyaloshinskii–Moriya (DM) interaction and  to an external uniform magnetic field. Intrinsic decoherence scheme offers and alternative to the study of coherence phenomena under symmetry breaking, just as Gong \textit{et. al} \cite{Gong2018} showed that in a ring arrangement of coupled harmonic oscillators. Guo-Hui and Bing Bing \cite{GuoHui2015} estimated the quantum discord of two qubits that loose coherence through  intrinsic mechanisms.  Furthermore, Mohamed \textit{et. al} \cite{Mohamed2020} have used the intrinsic decoherence effect for two qubits interacting with a coherent field, with the purpose to protect the entropy and entanglement from the dipole-dipole interaction. Le\'on-Montiel {\it et al.} \cite{LenMontiel2015}, have shown that noise may be helpful to assist energy transfer in coupled oscillators. Bayen and Mandal \cite{Bayen2020} reported the analytical solution of quartic anharmonic oscillator with driven force that show squeezing effects on the coherent states. some time ago, Mandal \cite{Mandal1998}, shows a way to calculate the quasidistribution of photons in the squeezed states of light when the parameters are taken as purely real. Germain \textit{et al.} \cite{Germain} explicitly shows the effects of a non-commutative bath of oscillators on a magneto-oscillator, they gives the solution for decoherence without dissipation using the master's equation formulation. Recently, Lu \cite{Lu2021}, reported the evolution of orthogonal coherent states given the solution of the master's equation, showing how the decoherence affects the negativity of the associate Wigner's function. Mohamed \cite{MOHAMED2013121} and Abhignan \cite{abhignan2021} have studied the role of intrinsic decoherence in the subsystems correlations by using the quantum discord, finding how the decoherence parameter leads to sudden death and revivals of some geometrical and dynamical quantities, like entanglement.

The present paper is organized as follow: In Section \ref{S1:Intrinsic} we give a review of the Milburn's intrinsic decoherence method. Section \ref{S2:DiHO} introduce the displaced harmonic oscillator in terms of annihilation and creation operators; there, we calculate the expectation values for the position quadrature and the number operators for a general initial condition. Subsection \ref{S2.3:CS} details the expectation values when we have an initial coherent state. Subsection \ref{S2.4:SS} details the expectation values when we have an initial squeezed state for complex squeezing parameter; here, we hold the real part constant and vary the phase to look for changes in the dynamics. Section \ref{S3:Concls} is left to discuss results and conclusions.

\section{Intrinsic decoherence equation and solution}\label{S1:Intrinsic}
Decoherence is a central research topic in quantum mechanics, since it shows how a quantum system starts to lose it quantumness because the system under study is usually surrounded by an environment that affects it. However, there may be other causes of decoherence that are intrinsic to the system \cite{Milburn1991}. We may say that there exist two main causes of decoherence \cite{Stamp2012}: First, those that are driven by environmental interactions, oscillators or spins bath, for example, where the Linblad master equations plays the role of the describing mathematical framework; and the other is intrinsic decoherence that is driven by the mere existence of the system dynamics in the equations that described their nature. We will study here the second,by using the equation that Milburn \cite{Milburn1991} introduced to modifiy the Schr\"odinger equation and that  describes (intrinsic) decoherence via the equation
\begin{equation}\label{0010}
    \dot{\rho}=\gamma \left(e^{-\mi\frac{\op{H}}{\gamma}}\rho e^{\mi\frac{\op{H}}{\gamma}}- \rho\right),
\end{equation}
where $\op{H}$ is the system's Hamiltonian, and $\gamma$ is the intrinsic decoherence parameter that states the rate of decaying in the dynamics. By developing the exponentials functions above in Taylor series, and keeping terms up to second order, we obtain
\begin{equation}\label{0011}
    \dot{\rho}\approx\gamma \left(\left[1-\mi\frac{\op{H}}{\gamma}-\frac{\op{H}^{2}}{\gamma^2}\right]\rho \left[1+\mi\frac{\op{H}}{\gamma}-\frac{\op{H}^2}{\gamma^2}\right]- \rho\right),
\end{equation}
that can be rewritten in the Lindblad form
\begin{equation*}
    \dot{\rho}=-\mi\left[\op{H},\rho\right]-\frac{1}{\gamma}\left[\op{H},\left[\op{H},\rho\right]\right],
\end{equation*}
where the Schr\"odinger equation is recovered when $\gamma\rightarrow\infty$. This last equation shows us that a Linbladian may be obtained from the complete intrinsic decoherence evolution, describing the system with different levels of accuracy, since the parameter expansion leads to drop highers order of decoherence.

However,  a complete solution of \eqref{0010} may be obtained
\begin{equation}\label{0020}
    \rho(t)=e^{-\gamma t}e^{\op{S}t}\rho(0),
\end{equation}
where the superoperator
\begin{equation*}
    \op{S}\rho=\gamma e^{-\mi\frac{\op{H}}{\gamma}}\rho e^{\mi\frac{\op{H}}{\gamma}},
\end{equation*}
has been used, such that
\begin{equation*}
    e^{\op{S}t}\rho(0)=\sum_{k=0}^{\infty}\frac{(\gamma t)^k}{k!}\, \rho_{k},
\end{equation*}
with the $k$-th element of the density matrix $\rho$ defined by
\begin{equation}\label{expt}
    \rho_{k}=\ket{\psi_{k}}\bra{\psi_k}, \qquad \ket{\psi_{k}}=e^{-\mi k\frac{\op{H}}{\gamma}}\ket{\psi(0)},
\end{equation}
with $\ket{\psi(0)}$  the initial wavefunction.

\section{Displaced harmonic oscillator}\label{S2:DiHO}
We start with the Hamiltonian for a displaced harmonic oscillator (we set $\hbar=1$)
\begin{equation}\label{hdis}
\op{H} = \omega \op{a}^{\dagger}\op{a}+\lambda\lpar{\op{a}+a^{\dagger}},
\end{equation}
where $\omega$ is the natural frequency of the standard harmonic oscillator, and $\lambda$ is the amplitude of the displacement, meaning that the potential function is moved and scaled.

It is not difficult to show that the Hamiltonian \eqref{hdis} may be rewritten as
\begin{equation}
\op{H} = \omega \op{D}^{\dagger}\lpar{\frac{\lambda}{\omega}}\op{a}^{\dagger}\op{a} \op{D}^{\dagger}\lpar{\frac{\lambda}{\omega}}-\frac{\lambda^2}{\omega},
\end{equation}
where $\op{D}\lpar{\tfrac{\lambda}{\omega}}$ is the usual displacement operator. With this last equation, we obtain the $k$-th element of the wavefunction \eqref{expt} as
\begin{equation}\label{Wfunc}
    \ket{\psi_{k}}=e^{\mi\frac{k\lambda^{2}}{\gamma\omega}}\op{D}^{\dagger}\left(\frac{\lambda}{\omega}\right)e^{-\mi\op{a}^{\dagger}\op{a}\frac{k\omega}{\gamma}}\op{D}\left(\frac{\lambda}{\omega}\right)\ket{\psi(0)}.
\end{equation}

We now study two properties of quantum mechanical system, namely the position quadrature and the average number of photons. It is interesting first, to denote how the decoherence parameter affects the position quadrature, since it represent the real part of the complex amplitude of the annihilation operator $\op{a}$; next, the effect on the photon number is explicit since we can observe how the population tends to decay in time depending on the strength of the decaying parameter $\gamma$.

\subsubsection*{Average of the position quadrature operator}
We start by calculating the matrix elements of (twice) the position quadrature operator over the $k$-th components of the wavefunction, we have then
\begin{equation}\label{psik_a}
\begin{aligned}
    &\bra{\psi_{k}}\op{a}^{\dagger}+\op{a}\ket{\psi_{k}} =\\ &\quad\lspar{\bra{\psi(0)}\op{D}^{\dagger}\lpar{\frac{\lambda}{\omega}}e^{\frac{\mi k \omega}{\gamma}\op{a}^{\dagger}\op{a}}\op{D}\lpar{\frac{\lambda}{\omega}}}\lpar{\op{a}^{\dagger}+\op{a}}\\
    &\qquad\qquad\times\lspar{\op{D}^{\dagger}\lpar{\frac{\lambda}{\omega}}e^{-\frac{\mi k \omega}{\gamma}\op{a}^{\dagger}\op{a}}\op{D}\lpar{\frac{\lambda}{\omega}}\ket{\psi(0)}}\\
    & = \bra{\psi(0)}\op{D}^{\dagger}\lpar{\frac{\lambda}{\omega}}e^{\frac{\mi k \omega}{\gamma}\op{a}^{\dagger}\op{a}}\lspar{\op{a}^{\dagger}+\op{a}-2\frac{\lambda}{\omega}}\\
    &\qquad\qquad\times e^{-\frac{\mi k \omega}{\gamma}\op{a}^{\dagger}\op{a}}\op{D}\lpar{\frac{\lambda}{\omega}}\ket{\psi(0)}\\
    & = \bra{\psi(0)}\op{D}^{\dagger}\lpar{\frac{\lambda}{\omega}}\lspar{\op{a}^{\dagger}e^{\frac{\mi k \omega}{\gamma}}+\op{a}e^{-\frac{\mi k \omega}{\gamma}}-2\frac{\lambda}{\omega}}\op{D}\lpar{\frac{\lambda}{\omega}}\ket{\psi(0)}\\
    & = \bra{\psi(0)}\lspar{\lpar{\op{a}^{\dagger}+\frac{\lambda}{\omega}}e^{\frac{\mi k \omega}{\gamma}}+\lpar{\op{a}+\frac{\lambda}{\omega}}e^{-\frac{\mi k \omega}{\gamma}}-2\frac{\lambda}{\omega}}\ket{\psi(0)},
\end{aligned}
\end{equation}
which gives finally
\begin{equation}\label{ann_average}
\begin{aligned}
    &\braket{(\op{a}^{\dagger}+\op{a})} = e^{-\gamma t}\sum\limits_{k = 0}^{\infty}\frac{\lpar{\gamma t}^{k}}{k!}\bra{\psi_{k}}\op{a}\ket{\psi_{k}}\\
    & = e^{-\gamma t}\sum\limits_{k = 0}^{\infty}\frac{\lpar{\gamma t}^{k}}{k!}\bra{\psi(0)}\left[\lpar{\op{a}^{\dagger}+\frac{\lambda}{\omega}}e^{\frac{\mi k \omega}{\gamma}}\right.\\
    &\hspace{0.37\columnwidth}\left.+\lpar{\op{a}+\frac{\lambda}{\omega}}e^{-\frac{\mi k \omega}{\gamma}}-2\frac{\lambda}{\omega}\right]\ket{\psi(0)}.
\end{aligned}
\end{equation}

Given a suitable initial wavefunction $\ket{\psi(0)}$, we can sum up the terms and find analytical expressions for the average dynamics.

\subsubsection*{Average of the number of photons operator}
For the average of the number of photons, we follow the procedure depicted in equations \eqref{psik_a} and \eqref{ann_average}, first calculating the matrix elements of the operator $\op{a}^{\dagger}\op{a}$
\begin{equation}
\begin{aligned}
    & \bra{\psi_{k}}\op{a}^{\dagger}\op{a}\ket{\psi_{k}} =\\
    & \quad\lspar{\bra{\psi(0)}\op{D}^{\dagger}\lpar{\frac{\lambda}{\omega}}e^{\frac{\mi k \omega}{\gamma}\op{a}^{\dagger}\op{a}}\op{D}\lpar{\frac{\lambda}{\omega}}}\op{a}^{\dagger}\op{a}\\
    & \hspace{2cm}\times\lspar{\op{D}^{\dagger}\lpar{\frac{\lambda}{\omega}}e^{-\frac{\mi k \omega}{\gamma}\op{a}^{\dagger}\op{a}}\op{D}\lpar{\frac{\lambda}{\omega}}\ket{\psi(0)}}\\
    & = \bra{\psi(0)}\op{D}^{\dagger}\lpar{\frac{\lambda}{\omega}}e^{\frac{\mi k \omega}{\gamma}\op{a}^{\dagger}\op{a}}\\
    & \hspace{2cm}\times \lspar{\op{a}^{\dagger}\op{a}-\frac{\lambda}{\omega}\lpar{\op{a}^{\dagger}+\op{a}}+\frac{\lambda^{2}}{\omega^{2}}}\\
    & \hspace{4cm}\times e^{-\frac{\mi k \omega}{\gamma}\op{a}^{\dagger}\op{a}}\op{D}\lpar{\frac{\lambda}{\omega}}\ket{\psi(0)}\\
    & = \bra{\psi(0)}\op{D}^{\dagger}\lpar{\frac{\lambda}{\omega}}\\
    & \hspace{2cm}\times\lspar{\op{a}^{\dagger}\op{a}-\frac{\lambda}{\omega}\lpar{\op{a}^{\dagger}e^{\frac{\mi k \omega}{\gamma}}+\op{a}e^{-\frac{\mi k \omega}{\gamma}}}+\frac{\lambda^{2}}{\omega^{2}}}\\
    & \hspace{6cm}\times\op{D}\lpar{\frac{\lambda}{\omega}}\ket{\psi(0)}\\
    & = \bra{\psi(0)}\left[\lpar{\op{a}^{\dagger}\op{a}+\frac{\lambda}{\omega}\lbr{\op{a}^{\dagger}+\op{a}}+2\frac{\lambda^{2}}{\omega^{2}}}-\right.\\ & \hspace{1.5cm}\left.\frac{\lambda}{\omega}\lpar{\lbr{\op{a}^{\dagger}+\frac{\lambda}{\omega}}e^{\frac{\mi k\omega}{\gamma}}+\lbr{\op{a}+\frac{\lambda}{\omega}}e^{-\frac{\mi k\omega}{\gamma}}}\right]\ket{\psi(0)},
\end{aligned}
\end{equation}
that in the same fashion of equation \eqref{ann_average}, we only need to give a suitable initial condition $\ket{\psi(0)}$ to obtain explicit expressions.

\subsection{Initial coherent state}\label{S2.3:CS}
Let $\ket{\psi(0)} = \ket{\alpha}$ be a coherent state, we can use this initial condition to evaluate the position quadrature and number operators averages in the way that equation \eqref{ann_average} dictates. 

\begin{figure}[htbp]
    \centering
    \includegraphics[width = 0.9\columnwidth, keepaspectratio]{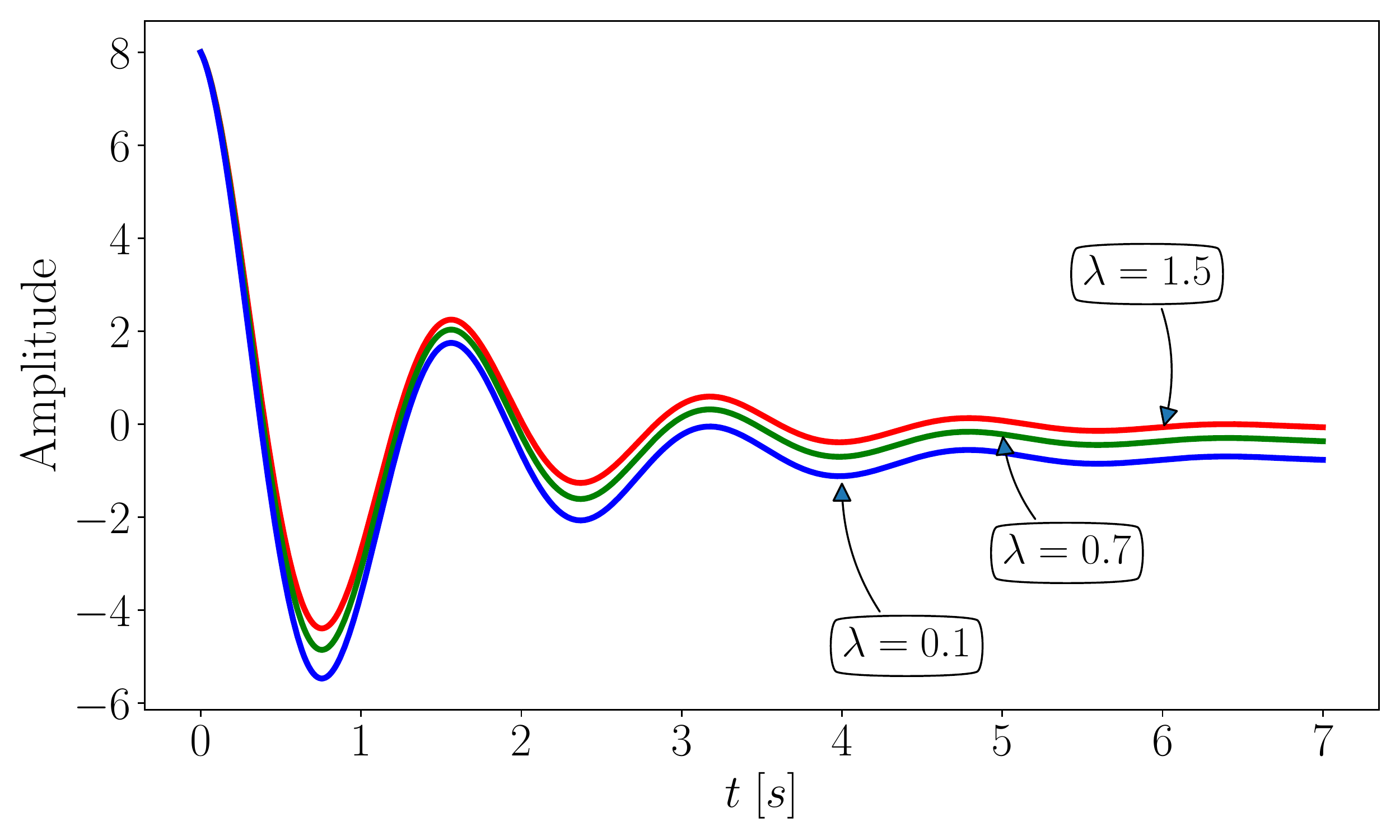}
    \caption{Expectation value of $\braket{(\op{a}^{\dagger}+\op{a})}$ given by \eqref{ann_avg}. The parameters are $\alpha = 4, \omega = 4, \gamma = 10$. We vary the displacement strength $\lambda\in\{0.1, 0.7, 1.5\}$.}
    \label{fig1}
\end{figure}

First, we start by calculating the dynamics of (twice) the position quadrature operator, since $\bra{\alpha}\op{a}^{\dagger}\ket{\alpha} = \alpha^{*}$ and $\bra{\alpha}\op{a}\ket{\alpha} = \alpha$, we obtain
\begin{equation}\label{ann_avg}
\begin{aligned}
    & \braket{\op{a}^{\dagger}+\op{a}} = e^{-\gamma t}\sum\limits_{k = 0}^{\infty}\frac{\lpar{\gamma t}^{k}}{k!}\bra{\psi_{k}}\op{a}^{\dagger}+\op{a}\ket{\psi_{k}}\\
    & = e^{-\gamma t}\sum\limits_{k = 0}^{\infty}\frac{\lpar{\gamma t}^{k}}{k!}\bra{\alpha}\\
    & \hspace{1.5cm}\times\lspar{\lpar{\op{a}^{\dagger}+\frac{\lambda}{\omega}}e^{\frac{\mi k \omega}{\gamma}}+\lpar{\op{a}+\frac{\lambda}{\omega}}e^{-\frac{\mi k \omega}{\gamma}}-2\frac{\lambda}{\omega}}\ket{\alpha}\\
    & = e^{-\gamma t}\sum\limits_{k = 0}^{\infty}\frac{\lpar{\gamma t}^{k}}{k!}\lspar{\lpar{\alpha^{*}+\frac{\lambda}{\omega}}e^{\frac{\mi k \omega}{\gamma}}+\lpar{\alpha+\frac{\lambda}{\omega}}e^{-\frac{\mi k \omega}{\gamma}}-2\frac{\lambda}{\omega}}\\
    & = e^{-\gamma t}\lspar{\lpar{\alpha^{*}+\frac{\lambda}{\omega}}e^{\gamma t e^{\frac{\mi \omega}{\gamma}}}+\lpar{\alpha+\frac{\lambda}{\omega}}e^{\gamma t e^{-\frac{\mi \omega}{\gamma}}}-2\frac{\lambda}{\omega}e^{\gamma t}}\\
    & = \lpar{\alpha^{*}+\frac{\lambda}{\omega}}e^{-\gamma t \lpar{1-e^{\frac{\mi \omega}{\gamma}}}}+\lpar{\alpha+\frac{\lambda}{\omega}}e^{-\gamma t \lpar{1-e^{-\frac{\mi \omega}{\gamma}}}}-2\frac{\lambda}{\omega},
\end{aligned}
\end{equation}
that checks $\braket{(\op{a}^{\dagger}+\op{a})} = \alpha^{*}+\alpha \equiv 2\mathrm{Re}\{\alpha\}$, when $t = 0$.
We plot the expectation value for $\braket{(\op{a}^{\dagger}+\op{a})}$ for several displacemet strengths in Fig. \ref{fig1}, showing the same behaviour, {\it i.e.}, a damping in the oscillations. Such damping is not amplified for larger displacement amplitudes.

\begin{figure}[htbp]
    \centering
    \includegraphics[width = 0.9\columnwidth, keepaspectratio]{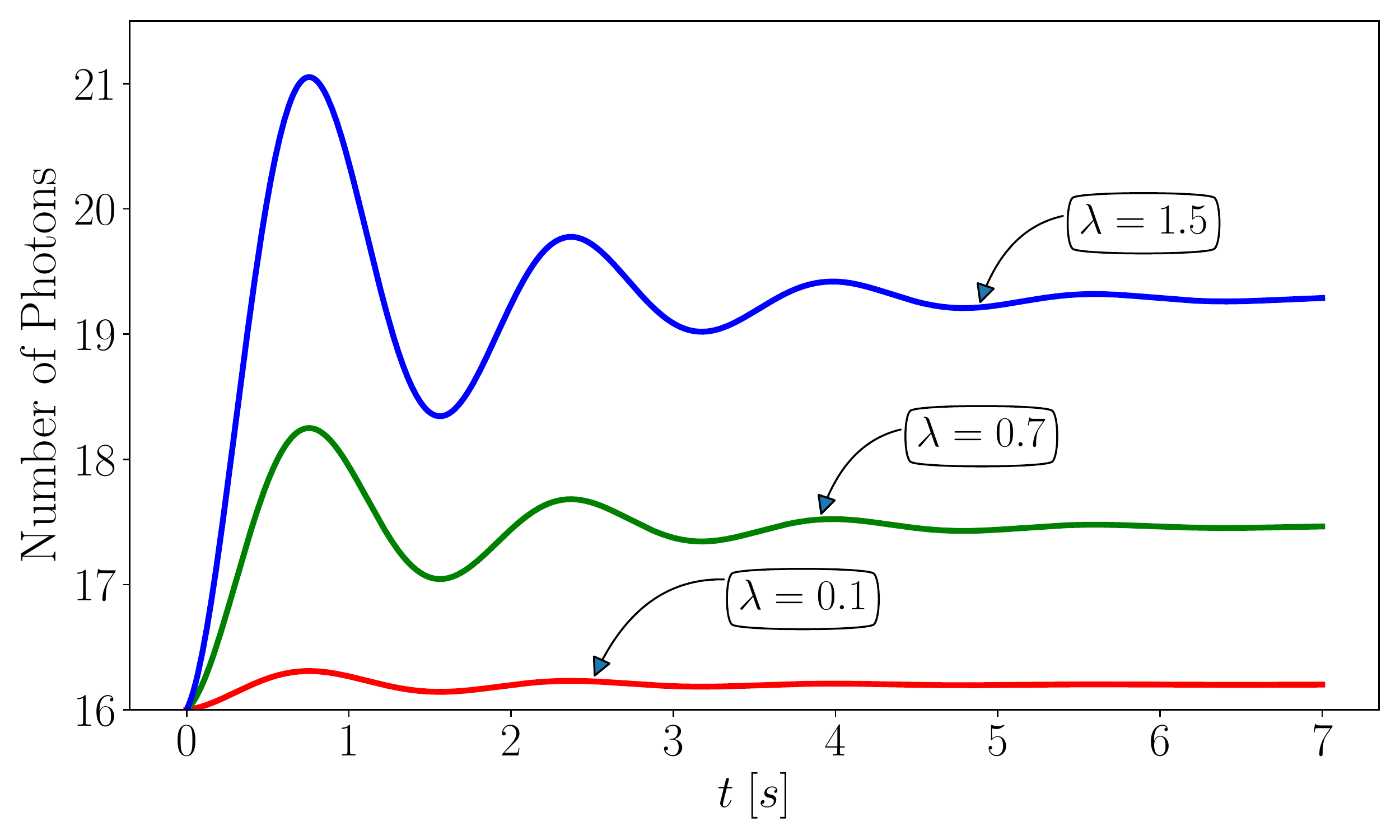}
    \caption{Expectation value of $\braket{\op{a}^{\dagger}\op{a}}$ given by \eqref{num_avg}. The parameters are $\alpha = 4, \omega = 4, \gamma = 10$. We variate the displacement strength $\lambda\in\{0.1, 0.7, 1.5\}$.}
    \label{fig2}
\end{figure}

For the average dynamics of the number of photons we have
\begin{equation}\label{num_avg}
\begin{aligned}
    & \braket{\op{a}^{\dagger}\op{a}} = e^{-\gamma t}\sum\limits_{k = 0}^{\infty}\frac{\lpar{\gamma t}^{k}}{k!}\bra{\psi_{k}}\op{a}^{\dagger}\op{a}\ket{\psi_{k}}\\
    & = e^{-\gamma t}\sum\limits_{k = 0}^{\infty}\frac{\lpar{\gamma t}^{k}}{k!}\bra{\alpha}\left[\lpar{\op{a}^{\dagger}\op{a}+\frac{\lambda}{\omega}\lbr{\op{a}^{\dagger}+\op{a}}+2\frac{\lambda^{2}}{\omega^{2}}}-\right.\\ & \hspace{2cm}\left.\frac{\lambda}{\omega}\lpar{\lbr{\op{a}^{\dagger}+\frac{\lambda}{\omega}}e^{\frac{\mi k\omega}{\gamma}}+\lbr{\op{a}+\frac{\lambda}{\omega}}e^{-\frac{\mi k\omega}{\gamma}}}\right]\ket{\alpha}\\
    & = e^{-\gamma t}\sum\limits_{k = 0}^{\infty}\frac{\lpar{\gamma t}^{k}}{k!}\left[\lpar{\left|\alpha\right|^{2}+\frac{\lambda}{\omega}\lbr{\alpha+\alpha^{*}}+2\frac{\lambda^{2}}{\omega^{2}}}-\right.\\ & \hspace{2cm}\left.\frac{\lambda}{\omega}\lpar{\lbr{\alpha+\frac{\lambda}{\omega}}e^{\frac{\mi k\omega}{\gamma}}+\lbr{\alpha^{*}+\frac{\lambda}{\omega}}e^{-\frac{\mi k\omega}{\gamma}}}\right]\\
    & = e^{-\gamma t}\left[\lpar{\left|\alpha\right|^{2}+\frac{\lambda}{\omega}\lbr{\alpha+\alpha^{*}}+2\frac{\lambda^{2}}{\omega^{2}}}e^{\gamma t}-\right.\\ & \hspace{2cm}\left.\frac{\lambda}{\omega}\lpar{\lbr{\alpha+\frac{\lambda}{\omega}}e^{\gamma t e^{\frac{\mi \omega}{\gamma}}}+\lbr{\alpha^{*}+\frac{\lambda}{\omega}}e^{\gamma t e^{-\frac{\mi \omega}{\gamma}}}}\right]\\
    & = \lspar{\left|\alpha\right|^{2}+\frac{\lambda}{\omega}\lbr{\alpha+\alpha^{*}}+2\frac{\lambda^{2}}{\omega^{2}}}-\\ & \hspace{1cm}\frac{\lambda}{\omega}\lspar{\lbr{\alpha+\frac{\lambda}{\omega}}e^{-\gamma t\lpar{1-e^{\frac{\mi \omega}{\gamma}}}}+\lbr{\alpha^{*}+\frac{\lambda}{\omega}}e^{-\gamma t\lpar{1-e^{-\frac{\mi \omega}{\gamma}}}}},
\end{aligned}
\end{equation}
that also checks $\braket{\op{a}^{\dagger}\op{a}}=\left|\alpha\right|^{2}$, for $t=0$.
In Fig. \ref{fig2} we plot the average number of photons which shows a similar behaviour as the quadrature, this is, a damping of the oscillations.

\subsection{Initial squeezed state}\label{S2.4:SS}
For an initial squeezed state $\ket{\alpha,z}=\op{S}(z)\ket{\alpha}$ we use that $\op{S}^{\dagger}(z)\op{a}\op{S}(z)=\mu\op{a} - \nu\op{a}^{\dagger}$ and $\op{S}^{\dagger}(z)\op{a}^{\dagger}\op{S}(z)=\mu\op{a}^{\dagger} - \nu^{*}\op{a}$, and because $z = r e^{\mi\theta}$, we have $\mu=\cosh r$ and $\nu=e^{\mi\theta}\sinh r$.

\begin{figure}[htbp]
    \centering
    \includegraphics[width = 0.9\columnwidth, keepaspectratio]{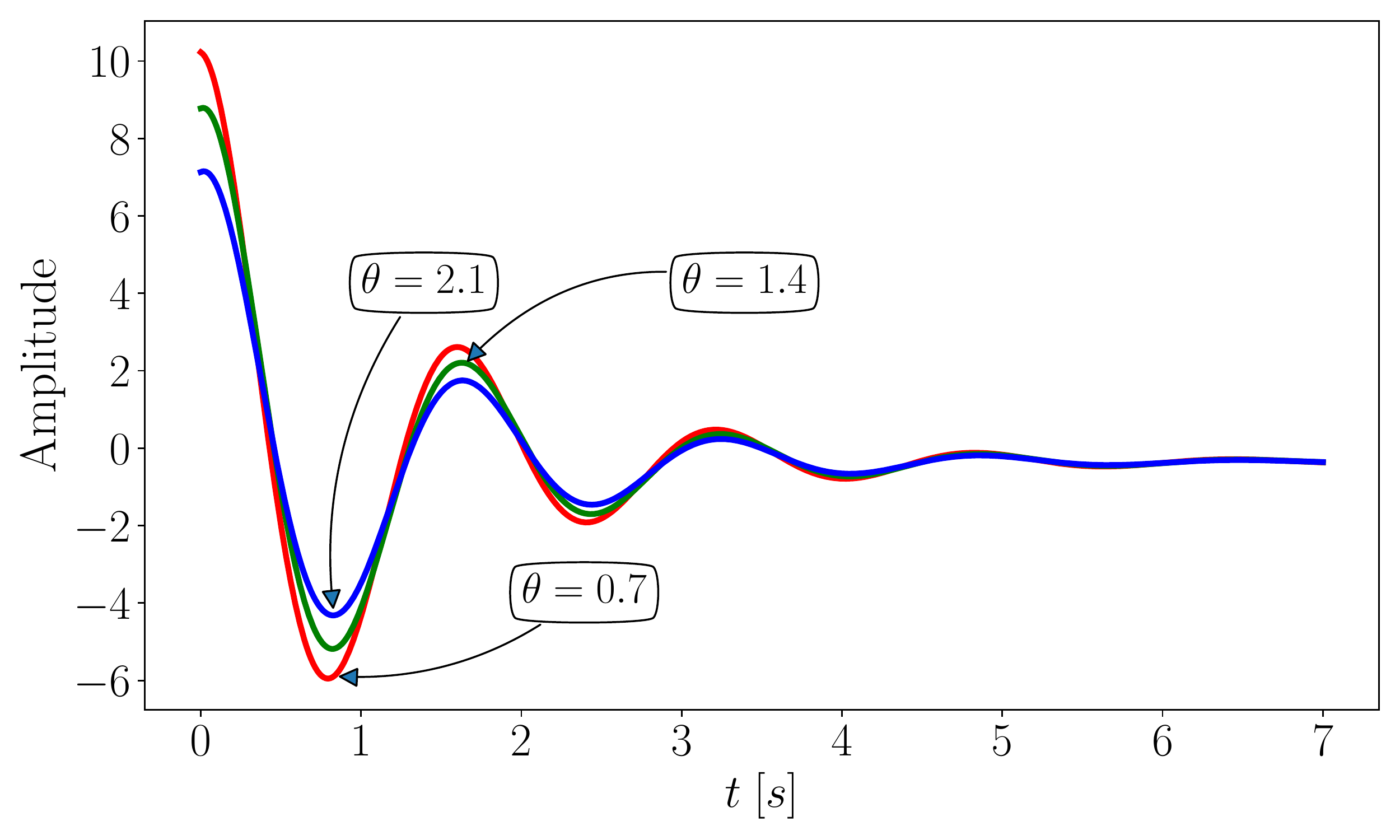}
    \caption{Expectation value of $\braket{(\op{a}^{\dagger}+\op{a})}$ given by \eqref{qsq_avg}. The parameters are $\alpha = 4, \omega = 4, \gamma = 10, \lambda = 0.7, r = 0.3$, where we choose to maintain the real part $r$ of the squeeze coefficient and variate their complex phase $\theta$.}
    \label{fig3}
\end{figure}

For the average of (twice) the position quadrature operator it is easy to show that the evolution gives
\begin{equation}\label{qsq_avg}
\begin{aligned}
    & \braket{\op{a}^{\dagger}+\op{a}} = e^{-\gamma t}\sum\limits_{k = 0}^{\infty}\frac{\lpar{\gamma t}^{k}}{k!}\bra{\psi_{k}}\op{a}^{\dagger}+\op{a}\ket{\psi_{k}}\\
    & = e^{-\gamma t}\sum\limits_{k = 0}^{\infty}\frac{\lpar{\gamma t}^{k}}{k!}\bra{\alpha,r}\left[\lpar{\op{a}^{\dagger}+\frac{\lambda}{\omega}}e^{\frac{\mi k \omega}{\gamma}}+\right.\\
    &\hspace{3cm}\left.\lpar{\op{a}+\frac{\lambda}{\omega}}e^{-\frac{\mi k \omega}{\gamma}}-2\frac{\lambda}{\omega}\right]\ket{\alpha,r}\\
    & = e^{-\gamma t}\sum\limits_{k = 0}^{\infty}\frac{\lpar{\gamma t}^{k}}{k!}\bra{\alpha}\op{S}^{\dagger}(r)\left[\lpar{\op{a}^{\dagger}+\frac{\lambda}{\omega}}e^{\frac{\mi k \omega}{\gamma}}+\right.\\
    & \hspace{3cm}\left.\lpar{\op{a}+\frac{\lambda}{\omega}}e^{-\frac{\mi k \omega}{\gamma}}-2\frac{\lambda}{\omega}\right]\op{S}(r)\ket{\alpha}\\
    & = e^{-\gamma t}\sum\limits_{k = 0}^{\infty}\frac{\lpar{\gamma t}^{k}}{k!}\left[\lpar{\mu\alpha^{*}-\nu^{*}\alpha+\frac{\lambda}{\omega}}e^{\frac{\mi k \omega}{\gamma}}+\right.\\
    &\hspace{3cm}\left.\lpar{\mu\alpha-\nu\alpha^{*}+\frac{\lambda}{\omega}}e^{-\frac{\mi k \omega}{\gamma}}-2\frac{\lambda}{\omega}\right]\\
    & = \lpar{\mu\alpha^{*}-\nu^{*}\alpha+\frac{\lambda}{\omega}}e^{-\gamma t \lpar{1-e^{\frac{\mi \omega}{\gamma}}}}+\\
    &\hspace{3cm}\lpar{\mu\alpha-\nu\alpha^{*}+\frac{\lambda}{\omega}}e^{-\gamma t \lpar{1-e^{-\frac{\mi \omega}{\gamma}}}}-2\frac{\lambda}{\omega}.
\end{aligned}
\end{equation}
Figure \ref{fig3} shows that, as in the case of the coherent states, the oscillations decay in a fast fashion, independently of the phase of the squeezed states.

For the average of the number operator, because the action of the squeezing operator results in
\begin{equation}
\begin{aligned}
    \op{S}^{\dagger}(z)\op{a}^{\dagger}\op{a}S(z) &= \lpar{\mu\op{a}^{\dagger}-\nu^{*}\op{a}}\lpar{\mu\op{a}-\nu\op{a}^{\dagger}}\\
    & = \mu^{2}\op{a}^{\dagger}\op{a}+|\nu|^{2}\op{a}\op{a}^{\dagger}-\mu\lpar{\nu\op{a}^{\dagger^2}+\nu^{*}\op{a}^{2}}\\
    & = \lpar{\mu^{2}+|\nu|^{2}}\op{a}^{\dagger}\op{a}-\mu\lpar{\nu\op{a}^{\dagger^2}+\nu^{*}\op{a}^{2}}+|\nu|^{2},
\end{aligned}
\end{equation}
we can then evaluate the average as

\begin{equation}\label{numsq_avg}
\begin{aligned}
    & \braket{\op{a}^{\dagger}\op{a}} = e^{-\gamma t}\sum\limits_{k = 0}^{\infty}\frac{\lpar{\gamma t}^{k}}{k!}\bra{\psi_{k}}\op{a}^{\dagger}\op{a}\ket{\psi_{k}}\\
    & = e^{-\gamma t}\sum\limits_{k = 0}^{\infty}\frac{\lpar{\gamma t}^{k}}{k!}\bra{\alpha,r}\left[\lpar{\op{a}^{\dagger}\op{a}+\frac{\lambda}{\omega}\lbr{\op{a}^{\dagger}+\op{a}}+2\frac{\lambda^{2}}{\omega^{2}}}-\right.\\ & \hspace{1cm}\left.\frac{\lambda}{\omega}\lpar{\lbr{\op{a}^{\dagger}+\frac{\lambda}{\omega}}e^{\frac{\mi k\omega}{\gamma}}+\lbr{\op{a}+\frac{\lambda}{\omega}}e^{-\frac{\mi k\omega}{\gamma}}}\right]\ket{\alpha,r}\\
    & = e^{-\gamma t}\sum\limits_{k = 0}^{\infty}\frac{\lpar{\gamma t}^{k}}{k!}\bra{\alpha}\op{S}^{\dagger}(r)\left[\lpar{\op{a}^{\dagger}\op{a}+\frac{\lambda}{\omega}\lbr{\op{a}^{\dagger}+\op{a}}+2\frac{\lambda^{2}}{\omega^{2}}}\right.\\ & \hspace{0.5cm}-\left.\frac{\lambda}{\omega}\lpar{\lbr{\op{a}^{\dagger}+\frac{\lambda}{\omega}}e^{\frac{\mi k\omega}{\gamma}}+\lbr{\op{a}+\frac{\lambda}{\omega}}e^{-\frac{\mi k\omega}{\gamma}}}\right]\op{S}(r)\ket{\alpha}\\
    & = e^{-\gamma t}\sum\limits_{k = 0}^{\infty}\frac{\lpar{\gamma t}^{k}}{k!}\left[\lbr{\mu^{2}+|\nu|^{2}}\left|\alpha\right|^{2}-\mu\lbr{\nu\alpha^{*^{2}}+\nu^{*}\alpha^{2}}+\right.\\
    &\hspace{1.5cm}\left.\frac{\lambda}{\omega}\lbr{\lbr{\mu-\nu^{*}}\alpha+\lbr{\mu-\nu}\alpha^{*}}+2\frac{\lambda^{2}}{\omega^{2}}+|\nu|^{2}-\right.\\
    & \left.\frac{\lambda}{\omega}\lpar{\lbr{\mu\alpha^{*}-\nu^{*}\alpha+\frac{\lambda}{\omega}}e^{\frac{\mi k\omega}{\gamma}}+\lbr{\mu\alpha-\nu\alpha^{*}+\frac{\lambda}{\omega}}e^{-\frac{\mi k\omega}{\gamma}}}\right]\\
    & = \left[\lbr{\mu^{2}+|\nu|^{2}}\left|\alpha\right|^{2}-\mu\lbr{\nu\alpha^{*^{2}}+\nu^{*}\alpha^{2}}+\right.\\ 
    & \hspace{1cm}\left.\frac{\lambda}{\omega}\lbr{\lbr{\mu-\nu^{*}}\alpha+\lbr{\mu-\nu}\alpha^{*}}+2\frac{\lambda^{2}}{\omega^{2}}+|\nu|^{2}\right]-\\
    & \hspace{2cm}\frac{\lambda}{\omega}\left[\lbr{\mu\alpha^{*}-\nu^{*}\alpha+\frac{\lambda}{\omega}}e^{-\gamma t\lpar{1-e^{\frac{\mi \omega}{\gamma}}}}+\right.\\
    & \hspace{3.5cm}\left.\lbr{\mu\alpha-\nu\alpha^{*}+\frac{\lambda}{\omega}}e^{-\gamma t\lpar{1-e^{-\frac{\mi \omega}{\gamma}}}}\right],
\end{aligned}
\end{equation}
\begin{figure}[htbp]
    \centering
    \includegraphics[width = 0.9\columnwidth, keepaspectratio]{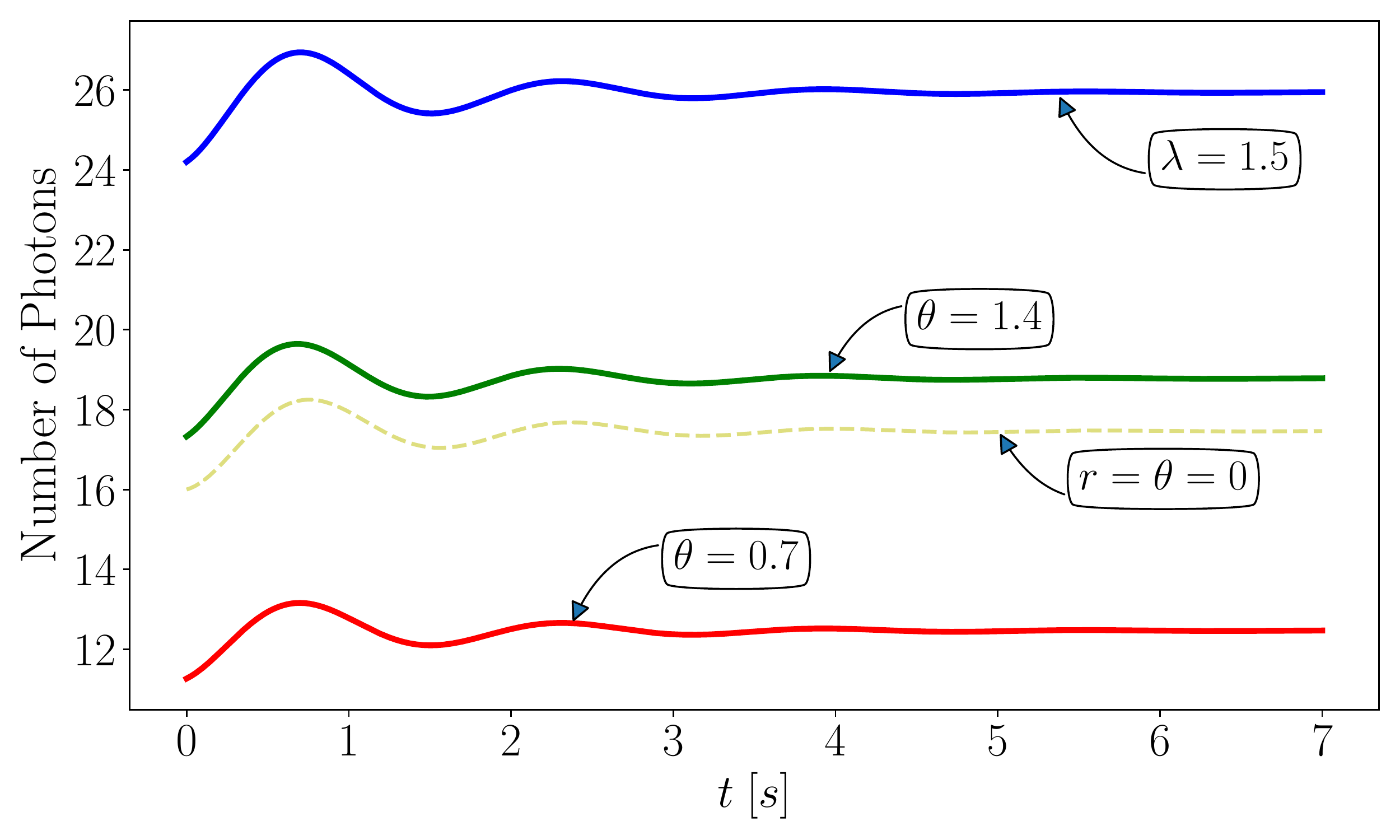}
    \caption{Expectation value of $\braket{\op{a}^{\dagger}\op{a}}$ given by \eqref{numsq_avg}. The parameters are $\alpha = 4, \omega = 4, \gamma = 10, \lambda = 0.7, r = 0.3$, where we choose to maintain the real part $r$ of the squeeze coefficient and variate their complex phase $\theta$.}
    \label{fig4}
\end{figure}
In Figure \ref{fig4} we plot the average number of photons which also shows, as former examples, the rapid decay of oscillations due to intrinsic decoherence.

\section{Conclusions}\label{S3:Concls}
We have given a  solution of the Milburn equation (beyond the Lindblad form that it is generally used). This equation describes a modification of the Scrh\"odinger equation that accounts for (intrinsic) decoherence. We have studied the decaying dynamics of a displaced harmonic oscillator for initially coherent and squeezed states.

\printbibliography

@article{Muthuganesan2021,
  doi = {10.1007/s11128-020-02985-y},
  url = {https://doi.org/10.1007/s11128-020-02985-y},
  year = {2021},
  month = jan,
  publisher = {Springer Science and Business Media {LLC}},
  volume = {20},
  number = {1},
  author = {R. Muthuganesan and V. K. Chandrasekar},
  title = {Intrinsic decoherence effects on measurement-induced nonlocality},
  journal = {Quantum Information Processing}
}

@article{Mohamed2021,
  doi = {10.1016/j.physe.2020.114529},
  url = {https://doi.org/10.1016/j.physe.2020.114529},
  year = {2021},
  month = apr,
  publisher = {Elsevier {BV}},
  volume = {128},
  pages = {114529},
  author = {A.-B.A. Mohamed and Abdel-Haleem Abdel-Aty and H. Eleuch},
  title = {Dynamics of trace distance and Bures correlations in a three-qubit {XY} chain: Intrinsic noise model},
  journal = {Physica E: Low-dimensional Systems and Nanostructures}
}

@article{He2021,
  doi = {10.1007/s10773-020-04693-w},
  url = {https://doi.org/10.1007/s10773-020-04693-w},
  year = {2021},
  month = jan,
  publisher = {Springer Science and Business Media {LLC}},
  volume = {60},
  number = {1},
  pages = {304--313},
  author = {Qi-Liang He and Min Ding and Yong-Jun Xiao and Xiao-Shu Song},
  title = {Quantum Coherence and Transfer of Quantum Information with a Kerr Medium Under Decoherence},
  journal = {International Journal of Theoretical Physics}
}

@article{Mohamed2020,
  doi = {10.1088/1402-4896/ab8f41},
  url = {https://doi.org/10.1088/1402-4896/ab8f41},
  year = {2020},
  month = may,
  publisher = {{IOP} Publishing},
  volume = {95},
  number = {7},
  pages = {075104},
  author = {A-B A Mohamed and H A Hessian and H Eleuch},
  title = {Generation of quantum coherence in two-qubit cavity system: qubit-dipole coupling and decoherence effects},
  journal = {Physica Scripta}
}

@article{Zheng2017,
  doi = {10.1140/epjd/e2017-80408-y},
  url = {https://doi.org/10.1140/epjd/e2017-80408-y},
  year = {2017},
  month = nov,
  publisher = {Springer Science and Business Media {LLC}},
  volume = {71},
  number = {11},
  author = {Li Zheng and Guo-Feng Zhang},
  title = {Intrinsic decoherence in Jaynes-Cummings model with Heisenberg exchange interaction},
  journal = {The European Physical Journal D}
}

@article{Yang2017,
  doi = {10.1088/1674-1056/26/1/010601},
  url = {https://doi.org/10.1088/1674-1056/26/1/010601},
  year = {2017},
  month = jan,
  publisher = {{IOP} Publishing},
  volume = {26},
  number = {1},
  pages = {010601},
  author = {Hong-ying Yang and Qiang Zheng and Qi-jun Zhi},
  title = {Optimal quantum parameter estimation of two-qutrit Heisenberg
		                    {XY}
		                    chain under decoherence},
  journal = {Chinese Physics B}
}

@article{MoyaCessa1993,
  doi = {10.1103/physreva.48.3900},
  url = {https://doi.org/10.1103/physreva.48.3900},
  year = {1993},
  month = nov,
  publisher = {American Physical Society ({APS})},
  volume = {48},
  number = {5},
  pages = {3900--3905},
  author = {H. Moya-Cessa and V. Bu{\v{z}}ek and M. S. Kim and P. L. Knight},
  title = {Intrinsic decoherence in the atom-field interaction},
  journal = {Physical Review A}
}

@article{Milburn1991,
  doi = {10.1103/physreva.44.5401},
  url = {https://doi.org/10.1103/physreva.44.5401},
  year = {1991},
  month = nov,
  publisher = {American Physical Society ({APS})},
  volume = {44},
  number = {9},
  pages = {5401--5406},
  author = {G. J. Milburn},
  title = {Intrinsic decoherence in quantum mechanics},
  journal = {Physical Review A}
}

@article{GuoHui2015,
  doi = {10.1007/s10773-015-2893-7},
  url = {https://doi.org/10.1007/s10773-015-2893-7},
  year = {2015},
  month = dec,
  publisher = {Springer Science and Business Media {LLC}},
  volume = {55},
  number = {5},
  pages = {2588--2597},
  author = {Yang Guo-Hui and Zhang Bing-Bing},
  title = {Quantum Discord Behaviors in Two Qubits Spin Squeezing Model with Intrinsic Decoherence},
  journal = {International Journal of Theoretical Physics}
}

@article{LenMontiel2015,
  doi = {10.1038/srep17339},
  url = {https://doi.org/10.1038/srep17339},
  year = {2015},
  month = nov,
  publisher = {Springer Science and Business Media {LLC}},
  volume = {5},
  number = {1},
  author = {Roberto de J. Le{\'{o}}n-Montiel and Mario A. Quiroz-Ju{\'{a}}rez and Rafael Quintero-Torres and Jorge L. Dom{\'{\i}}nguez-Ju{\'{a}}rez and H{\'{e}}ctor M. Moya-Cessa and Juan P. Torres and Jos{\'{e}} L. Arag{\'{o}}n},
  title = {Noise-assisted energy transport in electrical oscillator networks with off-diagonal dynamical disorder},
  journal = {Scientific Reports}
}

@article{Bayen2020,
  doi = {10.1140/epjp/s13360-020-00419-3},
  url = {https://doi.org/10.1140/epjp/s13360-020-00419-3},
  year = {2020},
  month = may,
  publisher = {Springer Science and Business Media {LLC}},
  volume = {135},
  number = {5},
  author = {Dolan Krishna Bayen and Swapan Mandal},
  title = {Squeezing of coherent light coupled to a periodically driven two-photon anharmonic oscillator},
  journal = {The European Physical Journal Plus}
}

@article{Mandal1998,
  doi = {10.1103/physreva.58.752},
  url = {https://doi.org/10.1103/physreva.58.752},
  year = {1998},
  month = jul,
  publisher = {American Physical Society ({APS})},
  volume = {58},
  number = {1},
  pages = {752--754},
  author = {Swapan Mandal},
  title = {Photon-number distribution of squeezed states:{\hspace{1em}}A graphical treatment},
  journal = {Physical Review A}
}

@article{Lu2021,
  doi = {10.1007/s12043-021-02169-y},
  url = {https://doi.org/10.1007/s12043-021-02169-y},
  year = {2021},
  month = aug,
  publisher = {Springer Science and Business Media {LLC}},
  volume = {95},
  number = {3},
  author = {Dao-Ming Lu},
  title = {Decoherence of orthogonal coherent state in the amplitude damping model},
  journal = {Pramana}
}

@article{Chlih2021,
   author = {Anas Ait Chlih and Nabil Habiballah and Mostafa Nassik},
   doi = {10.1007/S11128-021-03030-2},
   issn = {1573-1332},
   issue = {3},
   journal = {Quantum Information Processing 2021 20:3},
   month = {3},
   pages = {1-14},
   publisher = {Springer},
   title = {Dynamics of quantum correlations under intrinsic decoherence in a Heisenberg spin chain model with Dzyaloshinskii–Moriya interaction},
   volume = {20},
   url = {https://link.springer.com/article/10.1007/s11128-021-03030-2},
   year = {2021},
}

@misc{Germain,
  author       = {Yiande Deuto Germain and Azangue Koumetio Armel and Alain Giresse Tene and  Nsangou Isofa and Martin Tchoffo},
  doi = {10.1088/1402-4896/ac0273},
   issue = {8},
   journal = {Physica Scripta},
   month = {7},
   pages = {085705},
   publisher = {IOP},
   title = {Decoherence dynamics of a charged particle within a non-demolition type interaction in non-commutative phase-space},
   volume = {96},
   url = {https://doi.org/10.1088/1402-4896/ac0273},
   year = {2021},
}

@article{Gong2018,
   author = {ZhiRui Gong and ZhenWei Zhang and DaZhi Xu and Nan Zhao and ChangPu Sun},
   doi = {10.1007/S11433-017-9101-4},
   issn = {1869-1927},
   issue = {4},
   journal = {Science China Physics, Mechanics \& Astronomy 2018 61:4},
   month = {1},
   pages = {1-13},
   publisher = {Springer},
   title = {Spontaneous decoherence of coupled harmonic oscillators confined in a ring},
   volume = {61},
   url = {https://link.springer.com/article/10.1007/s11433-017-9101-4},
   year = {2018},
}

@article{MOHAMED2013121,
title = {Geometric Measure of Nonlocality and Quantum Discord of Two Charge Qubits with Phase Decoherence and Dipole-Dipole Interaction},
journal = {Reports on Mathematical Physics},
volume = {72},
number = {1},
pages = {121-132},
year = {2013},
issn = {0034-4877},
doi = {https://doi.org/10.1016/S0034-4877(14)60009-4},
url = {https://www.sciencedirect.com/science/article/pii/S0034487714600094},
author = {Abdel-Baset A. Mohamed}
}

@article{abhignan2021,
  doi = {10.1088/1402-4896/ac322f},
  url = {https://doi.org/10.1088/1402-4896/ac322f},
  year = {2021},
  month = nov,
  publisher = {{IOP} Publishing},
  volume = {96},
  number = {12},
  pages = {125114},
  author = {Venkat Abhignan and R Muthuganesan},
  title = {Effects of intrinsic decoherence on discord-like correlation measures of two-qubit spin squeezing model}
}

@article{Stamp2012,
author = {Stamp, P. C. E. },
title = {Environmental decoherence versus intrinsic decoherence},
journal = {Philosophical Transactions of the Royal Society A: Mathematical, Physical and Engineering Sciences},
volume = {370},
number = {1975},
pages = {4429-4453},
year = {2012},
doi = {10.1098/rsta.2012.0162},

URL = {https://royalsocietypublishing.org/doi/abs/10.1098/rsta.2012.0162},
eprint = {https://royalsocietypublishing.org/doi/pdf/10.1098/rsta.2012.0162}
}

\end{document}